\documentclass{CUP-JNL-DTM}%

\usepackage{graphicx}
\usepackage{multicol,multirow}
\usepackage{amsmath,amssymb,amsfonts}
\usepackage{mathrsfs}
\usepackage{amsthm}
\usepackage{rotating}
\usepackage{appendix}
\usepackage[numbers]{natbib}
\usepackage{ifpdf}
\usepackage[T1]{fontenc}
\usepackage{newtxtext}
\usepackage{newtxmath}
\usepackage{textcomp}
\usepackage{xcolor}
\usepackage{lipsum}
\usepackage[colorlinks,allcolors=blue]{hyperref}

\usepackage{placeins}

\usepackage{mhchem}
\usepackage[nolist,nohyperlinks]{acronym}
\usepackage{enumerate}

\theoremstyle{definition}

\numberwithin{equation}{section}

\jname{Data-Centric Engineering}
\articletype{RESEARCH ARTICLE}

\jyear{YEAR}

\begin{document}

\begin{acronym}
\acro{DNS}{Direct Numerical Simulation}
\acro{LES}{Large Eddy Simulation}
\acro{CFD}{Computational Fluid Dynamics}
\acro{MSE}{Mean Squared Error}
\acro{RMSE}{Root Mean Square Error}
\acro{SGS}{Subgrid-Scale}
\acro{HRR}{Heat Release Rate}
\acro{NN}{Neural Network}
\acro{HPC}{High Performance Computing}
\acro{FSD}{Flame Surface Density}
\acro{ML}{Machine Learning}
\acro{DL}{Deep Learning}
\acro{AI}{Artificial Intelligence}
\acro{NN}{Neural Network}
\acro{CNN}{Convolutional Neural Network}
\acro{TF}{Thickened Flame}
\acro{LR}{Learning Rate}
\acro{1D}{One-Dimensional}
\acro{2D}{Two-Dimensional}
\acro{3D}{Three-Dimensional}
\acro{NSCBC}{Navier-Stokes Characteristic Boundary Condition}
\acro{DDP}{Distributed Data Parallel}
\acro{PDF}{Probability Density Function}
\end{acronym}

\begin{Frontmatter}

\title{Hydrogen reaction rate modeling based on convolutional neural network for large eddy simulation}

\author[1]{Quentin Malé}
\author[2]{Corentin J. Lapeyre}
\author[1]{Nicolas Noiray}

\authormark{Quentin Malé \textit{et al}.}

\address[1]{CAPS Laboratory, Department of Mechanical and Process Engineering, ETH Zürich, Zürich 8092, Switzerland}

\address[2]{NVIDIA Corporation, Santa Clara, CA, United States}

\authormark{Quentin Malé et al.}

\keywords{data-driven reacting flow modeling;
hydrogen combustion;
machine learning;
artificial intelligence;
computational fluid dynamics}

\makeatletter
\g@addto@macro\@keywords{%
  \par\addvspace{10pt}%
  \noindent\fbox{%
    \parbox{0.6\textwidth}{\small
      This is a pre-print of an article published in:\\
      \textbf{Data-Centric Engineering}.\\
      The journal version is available online at:\\
      \href{https://doi.org/10.1017/dce.2025.1}{\textbf{https://doi.org/10.1017/dce.2025.1}}\\
      Please cite the journal version rather than the arXiv entry.%
    }%
  }\par
}
\makeatother

\abstract{
This paper establishes a data-driven modeling framework for lean Hydrogen (H\textsubscript{2})-air reaction rates for the Large Eddy Simulation (LES) of turbulent reactive flows. 
This is particularly challenging since H\textsubscript{2} molecules diffuse much faster than heat, leading to large variations in burning rates, thermodiffusive instabilities at the subfilter scale, and complex turbulence-chemistry interactions. 
Our data-driven approach leverages a Convolutional Neural Network (CNN), trained to approximate filtered burning rates from emulated LES data. 
First, five different lean premixed turbulent H\textsubscript{2}-air flame Direct Numerical Simulations (DNSs) are computed each with a unique global equivalence ratio. Second, DNS snapshots are filtered and downsampled to emulate LES data. Third, a CNN is trained to approximate the filtered burning rates as a function of LES scalar quantities: progress variable, local equivalence ratio and flame thickening due to filtering. 
Finally, the performances of the CNN model are assessed on test solutions never seen during training. The model retrieves burning rates with very high accuracy. 
It is also tested on two filter and downsampling parameters and two global equivalence ratios between those used during training. 
For these interpolation cases, the model approximates burning rates with low error even though the cases were not included in the training dataset. 
This a priori study shows that the proposed data-driven machine learning framework is able to address the challenge of modeling lean premixed H\textsubscript{2}-air burning rates. It paves the way for a new modeling paradigm for the simulation of carbon-free hydrogen combustion systems.
}

\end{Frontmatter}

\section*{Impact Statement}
Large Eddy Simulation (LES) plays a key role in the design of new combustion chambers for robust and safe combustion of carbon-free fuels such as H\textsubscript{2}. It is a very efficient method which consists of filtering and modeling small-scale flow physics that would otherwise be very computationally expensive to resolve. 
However, the high molecular diffusion of H\textsubscript{2} challenges current subfilter-scale modeling approaches for lean premixed mixtures. 
By establishing a framework for data-driven machine learning modeling of LES burning rates, this work can enable high-fidelity simulation of hydrogen combustion systems. These combustion systems are essential for achieving a low-carbon economy.


%
\section{Introduction}

The urgent need to drastically reduce our greenhouse gas emissions requires us to develop alternative solutions to fossil fuel combustion. Sustainably produced hydrogen (H\textsubscript{2}) is a promising energy carrier for the decarbonization of combustion systems where there is no viable technology to be a substitute. This is the case, for example, with gas turbines for power generation or aviation, as well as some industrial burners. 
However, compared with conventional fuels, H\textsubscript{2}-air combustion involves higher flame temperatures, much higher flame speeds, and a very wide flammability range. These characteristics pose considerable challenges for the design of new H\textsubscript{2} combustion devices, ensuring stable, robust and safe operation. In this context, high-fidelity simulation of turbulent combustion plays a key role. It is essential both for a fundamental understanding of flame behavior, and as an aid to the design of industrial combustion devices. It is also crucial for modeling safety scenarios, with hydrogen expected to be widely used to decarbonize industry.

The community nowadays uses \ac{LES} methods, which involve simulating only the largest flow structures, to compute real complex configurations at an affordable computing cost. The smaller scales are modeled, according to subfilter-scale models \cite{poinsot_theoretical_2011,fiorina_modeling_2015}. In LES, a filtered quantity $\varphi$ is defined as
\begin{equation}
    \overline{\varphi}(\mathbf{x},t) = \int_{-\infty}^\infty \varphi \left(\mathbf{x}^{\prime},t\right) F\left(\mathbf{x}-\mathbf{x}^{\prime}\right) d \mathbf{x}^{\prime} \, \text{,}
    \label{eq:filtering}
\end{equation}
where $F$ is the \ac{LES} filter. Balance equations are obtained by applying the filter $F$ to the multispecies Navier-Stokes equations. The filtered transport equation of a reactive scalar $\varphi$ reads
\begin{equation}
    \partial_t \left(\overline{\rho} \widetilde{\varphi}\right) + \nabla \cdot \left(\overline{\rho} \widetilde{\boldsymbol{u}} \widetilde{\varphi}\right) - \nabla \cdot ( \overline{\rho} \widetilde{D}_\varphi \nabla \widetilde{\varphi} ) = \nabla \cdot \Gamma_\varphi + \overline{\dot{\omega}}_\varphi \text{ ,}
    \label{eq:LES_transport_varphi}
\end{equation}
where $\widetilde{\cdot}$ denotes density-weighted Favre filtering (i.e. $\widetilde{\cdot} = \overline{\rho \cdot}/\overline{\rho}$), with density $\rho$, time $t$, velocity $\boldsymbol{u}$. $D_\varphi$, $\Gamma_\varphi$ and ${\dot{\omega}}_\varphi$ are the molecular diffusivity, subfilter-scale flux and source term of $\varphi$. The filtering operation produces unclosed terms located on the RHS of Eq.~(\ref{eq:LES_transport_varphi}) that must be modeled. 
The subfilter-scale flux comprises the transport of $\varphi$ by unresolved fluctuations of molecular diffusive flux and momentum
\begin{equation}
    \Gamma_\varphi = \Gamma_{\varphi, D} - \Gamma_{\varphi, C} \; \; \text{with} \; \Gamma_{\varphi, D} = \overline{\rho D_\varphi \nabla \varphi} - \overline{\rho} \widetilde{D}_\varphi \nabla \widetilde{\varphi} \; \; \text{and} \; \Gamma_{\varphi, C} = \overline{\rho \mathbf{u} \varphi }-\overline{\rho} \widetilde{\mathbf{u}} \widetilde{\varphi} \; \text{.}
\end{equation}
It is usually modeled neglecting $\Gamma_{\varphi, D}$ and involving gradient transport models for $\Gamma_{\varphi, C}$ with a turbulent viscosity hypothesis \cite{boger_direct_1998}. Countergradient phenomenon can be taken into account via more complex modeling \cite{moreau_counter-gradient_2002,tullis_scalar_2002}, although a large part will be described at the resolved scale in \ac{LES} \cite{boger_direct_1998}. Alternatively, $\Gamma_\varphi$ can be tabulated within the filtered tabulated chemistry framework \cite{fiorina_filtered_2010} or modeled via \ac{ML} algorithms \cite{seltz_direct_2019}. 
Modeling of the filtered chemical source term $\overline{{\dot{\omega}}}_\varphi$ is extremely challenging because combustion generally occurs within reaction fronts that are smaller than the LES computational mesh. Furthermore, there is a strong interaction between turbulence and chemistry: chemical heat release influences turbulence, while turbulence, in turn, modifies flame structure, which can either enhance or inhibit chemical reactions. 

To overcome these difficulties, several approaches have been proposed in the context of premixed combustion, which is the focus of this paper. A common technique is to artificially thicken the reaction front, enabling it to be resolved on the given computational mesh. 
This is the basis of the \ac{TF} model, which achieves this by increasing all diffusivities by a factor $F$ while decreasing all reaction rates by the same factor to maintain the correct flame propagation speeds \cite{butler_numerical_1977}. The LES modeling then focuses on an efficiency function $E$ taking into account the effects of thickening on flame propagation, such as the loss of flame wrinkling \cite{colin_thickened_2000,charlette_power-law_2002,poinsot_theoretical_2011}. 
Another approach consists of filtering the flame front with a LES filter larger than the mesh size \cite{boger_direct_1998,duwig_study_2007}. Filtering is usually done on the equation of a reaction progress variable $c$ representing the evolution of combustion within the flame front. 
The balance equation for $\widetilde{c}$ may be recast as a level set G-equation \cite{kerstein_field_1988}
\begin{equation}
    \partial_t \left( \bar{\rho} \widetilde{c} \right) + \nabla \cdot \left( \bar{\rho} \widetilde{\boldsymbol{u}} \widetilde{c} \right) + \nabla \cdot \left( \overline{\rho \boldsymbol{u} c} - \overline{\rho} \widetilde{\boldsymbol{u}} \widetilde{c} \right) = \nabla ( \overline{\rho D_c \nabla \cdot c} ) + \overline{\dot{\omega}}_c = \overline{\rho s_d |\nabla c|} \; \text{,}
    \label{eq:FPV}
\end{equation}
which gives rise to a flame front displacement term at speed $s_d$, $\overline{\rho s_d |\nabla c|}$, which requires modeling. 
Another way to close the equation is to tabulate all the closure terms from filtered laminar premixed flames \cite{fiorina_filtered_2010}, assuming the turbulent flame can be represented as an ensemble of locally \ac{1D} laminar flamelets \cite{peters_laminar_1988}. 
In all cases, a model to retrieve the subfilter-scale effects on the thickened/filtered flame front is needed.

Modeling subfilter-scale phenomena is particularly challenging for hydrogen combustion. Indeed, H\textsubscript{2} is a very peculiar fuel due to its small Lewis number. Its high molecular diffusion, much faster than heat, gives rise to unique phenomena, especially when burned lean (i.e. with excess air) and premixed. First, lean H\textsubscript{2} flames are prone to thermodiffusive instabilities. The disparity between heat and H\textsubscript{2} mass fluxes (called differential diffusion) amplifies small flame front perturbations, inducing significant flame wrinkling and strong variations of the reaction rates along the flame front \cite{poinsot_theoretical_2011,berger_intrinsic_2022,berger_intrinsic_2022-1}. 
Second, burning rates increase significantly when lean H\textsubscript{2} flame is subjected to a strain \cite{petrov_unsteady_1994,Vagelopoulos_further_1994}, up to a certain limit. The resulting increase in flame speed can be substantial. For example, the experimental work in Ref.~\cite{Vagelopoulos_further_1994} reports a four-fold enhancement in flame speed compared to an unstrained flame. 

These phenomena invalidate usual \ac{FSD} approaches \cite{pope_evolution_1988,marble_coherent_1997,boger_direct_1998,trouve_evolution_1994} based on the assumption that subfilter-scale effects are limited to an increase in flame surface density through wrinkling due to turbulence alone. 
Aniello et al. \cite{aniello_introducing_2022} have recently attempted to address this issue by multiplying the filtered chemical source terms by a corrective factor, determined from \ac{DNS} \cite{berger_intrinsic_2022,berger_intrinsic_2022-1}, decoupling the enhancement of combustion rates due to thermodiffusive instabilities and to usual turbulence wrinkling. However, this is no longer possible when the wrinkling scales induced by turbulent eddies become smaller than the characteristic lengths at which the instabilities occur. Moreover, Berger et al. \cite{berger_synergistic_2022} have reported complex and synergistic interactions between thermodiffusive instabilities and turbulence, casting doubt on the possibility of decoupling these phenomena. 
Gaucherand et al. \cite{gaucherand_subgrid-scale_2024} attempted to correct for unresolved straining effects by multiplying the filtered chemical source terms by a strain-to-unstrained flame speed ratio determined using \ac{1D} flames. The strain rate was assumed to be constant over the entire domain at a value of $2000$ $\mathrm{s^{-1}}$, not taking into account the very wide strain variation along the turbulent flame. Additional efforts must be made to account for local strain effects, which is extremely difficult in LES because the information on the strain that a turbulent flame actually undergoes at the subfilter-scale level is unknown. 
Subfilter \ac{PDF} modeling for H\textsubscript{2} turbulent combustion is also being developed \cite{berger_development_2022,pitsch_transition_2024,yao_capturing_2024,ferrante_differential_2024}, where the subfilter-scale distribution of thermochemical variables is usually modeled using $\beta$-PDF. The turbulence-chemistry interactions are described through the subfilter \ac{PDF} while the flamelet assumption \cite{peters_laminar_1988} links local thermochemical variables to those of the laminar flame elements. Thus, a set of \ac{1D} laminar flames is needed to be computed, assuming a specific flame archetype a priori. 
Defining the subfilter PDF is a particular challenge. For example, Refs~\cite{berger_development_2022,yao_capturing_2024} use a Dirac function for the mixture fraction distribution, while it is reported in Ref.~\cite{pitsch_transition_2024} that it leads to modeling errors. Indeed, preferential diffusion in H\textsubscript{2}-air mixtures can lead to fluctuation of the mixture fraction at the subfilter-scale level. Furthermore, the effect of unstable modes that occur partially at the subfilter-scale must be included in the subfilter PDF to avoid underestimating the flame consumption speed \cite{lapenna_-posteriori_2024,pitsch_transition_2024}. 

The phenomena involved in H\textsubscript{2} turbulent combustion being extremely complex, not well understood, and still subject to discussion, we propose to change the paradigm by developing a data-driven method based on \ac{ML} instead of trying to model turbulent H\textsubscript{2} burning rates by physical description. This work is motivated by the positive results of such an approach for the approximation of the subfilter flame surface density both a priori and a posteriori for methane-air combustion \cite{lapeyre_training_2019,lapeyre_-posteriori_2018,ho_augmenting_2024}. The strategy is adapted to hydrogen combustion by training a \ac{NN} to estimate the H\textsubscript{2} burning rate directly. In this way, no assumptions are made about the turbulent combustion regime, and the model can theoretically be applied to any combustion regime encountered during training. Firstly, a database of fully resolved turbulent H\textsubscript{2}-air premixed flames is generated using \ac{DNS}, taking into account preferential and differential diffusion effects. The database includes five different equivalence ratios to assess the ability of the model to account for air/fuel ratio variation. The database is then filtered and downsampled to emulate data from \ac{LES}. Three different filter sizes are used to assess the ability of the model to account for subfilter effects at different scales. From the emulated \ac{LES} fields, a \ac{CNN} is trained to estimate the LES burning rates based on known LES scalar quantities such as the filtered mixture fraction and progress variable. Once trained, the model is found to approximate the burning rates on solutions never seen during training with high accuracy. In addition to this, the model is able to approximate burning rates on filter sizes and equivalence ratios other than those used for training.

First, the problem statement is described, with the definition of the input/output variables of the model (Section~\ref{sec:pb_statement}). Then the \ac{NN} set up to approximate the burning rate function is detailed (Section~\ref{sec:ML_framework}). Next, the data generation and training strategy is explained (Section~\ref{sec:data_training_strategy}). Finally, training results are presented and discussed (Section~\ref{sec:results}).

%
\section{Background and methodology}

\subsection{Problem statement}\label{sec:pb_statement}

The simulation of turbulent flows at high Reynolds numbers cannot generally be performed with sufficient resolution to represent all the scales of the flow. \ac{LES} aims to simulate only a part of the turbulent spectrum to achieve a reasonable computational cost. Scales are cut by spatial filtering, and unresolved scales are modeled using so-called subfilter-scale models. In the LES framework, the Favre-filtered transport equation for the mass fraction of the $k^\text{th}$ species $Y_k$ follows Eq.~(\ref{eq:LES_transport_varphi}). 
As discussed in the introduction, modeling the filtered chemical source terms in LES of reacting flows is extremely challenging because chemistry largely occurs at the subfilter-scale level in most cases. This is further complicated in lean premixed H\textsubscript{2}-air mixtures because of thermodiffusive effects, causing additional flame wrinkling and pronounced fluctuations in reaction rates along the flame front. A precise description of the preferential and differential diffusion, turbulence wrinkling, and their interactions \cite{berger_synergistic_2022} at subfilter-scale level is required. 

In this work, we focus on the development of a data-driven model for the filtered chemical source term of H\textsubscript{2} in lean premixed turbulent combustion. \ac{ML} is used to discover structural relations between resolved scalar quantities and subfilter-scale turbulence-chemistry interactions. We develop this method for the H\textsubscript{2} filtered ``burning rate'' $\overline{\dot{\omega}}$, defined as 
\begin{equation}
    \overline{\dot{\omega}} = - \overline{\dot{\omega}}_\mathrm{H_2} \text{ ,}
    \label{eq:omega_bar}
\end{equation}
with the aim of establishing an \ac{ML}-based modeling framework and a proof of concept for H\textsubscript{2} turbulent combustion. Final implementation will depend on the given \ac{LES} modeling framework. For example, the \ac{ML} model could replace the tabulation method proposed in Refs~\cite{lapenna_data-driven_2021,lapenna_-posteriori_2024} for the chemical source term of the progress variable. This is what has been done in Ref.~\cite{seltz_direct_2019} with direct mapping of the closure terms for a methane-air flame. The ML framework could also be based on a database with one-step chemistry \cite{millan-merino_new_2024,schiavone_arrhenius-based_2024} to close the LES multi-species transport equations. 

The basic ingredients to describe turbulent flames remain the quantities introduced for laminar flame analysis: the progress variable $c$ and the mixture fraction $\xi$. The progress variable equals zero in fresh gas, one in burnt gas, and varies monotonically during combustion. Mixture fraction $\xi$ or the equivalence ratio $\phi$ is used to inform the \ac{ML}-based model of the composition of the mixture, which impacts burning rates. The mixture fraction is based on Bilger's definition \cite{bilger_reduced_1990}
\begin{equation}
    {\xi} = \frac{{\beta} - \beta^o}{\beta^f - \beta^o} \; \text{with} \; \beta = \frac{Z_\mathrm{H}}{2 W_\mathrm{H}} - \frac{Z_\mathrm{O}}{W_\mathrm{O}} \, \text{,}
    \label{eq:xi}
\end{equation}
where $Z_m$ and $W_m$ are the elemental mass fraction and atomic mass of the element $m$. The superscripts $o$ and $f$ denote oxidizer and fuel stream. 
The filtered progress variable $\widetilde{c}$ is defined as
\begin{equation}
    \widetilde{c} = \frac{Y_\mathrm{H_2}^u(\widetilde{\xi}) - \widetilde{Y}_\mathrm{H_2}}{Y_\mathrm{H_2}^u(\widetilde{\xi}) - Y_\mathrm{H_2}^b(\widetilde{\xi})} = \frac{\widetilde{\xi} - \widetilde{Y}_\mathrm{H_2}}{\widetilde{\xi} - \mathrm{max}\left(0, \frac{\widetilde{\xi}-\xi_{s}}{1-\xi_s}\right)} \, \text{,}
    \label{eq:c_tilde}
\end{equation}
where superscripts $u$ and $b$ denote values in unburnt and burnt gas and $\xi_s$ is the mixture fraction value at stoichiometry. We have arbitrarily chosen to base the progress variable on the H\textsubscript{2} mass fraction. It is shown in supplementary material that it is also possible to base the progress variable on the H\textsubscript{2}O mass fraction without any impact on the training and prediction accuracy of the model. In this framework, the challenge is now to model $\overline{\dot{\omega}}$ as a function of $\widetilde{c}$ and $\widetilde{\xi}$ fields.

\subsection{Machine learning framework}\label{sec:ML_framework}

The aim of the \ac{ML}-based model is to approximate the function 
\begin{equation}
    \overline{\dot{\omega}} = f \left( \widetilde{c}, \widetilde{\xi} \, \text{or} \, \widetilde{\phi}, \ldots \right) \, \text{.}
\end{equation}
The evolution of the reaction rate depends on the thermochemical state described by progress variable and mixture fraction, but also on the flow. Indeed, the level of turbulence, flame curvature and strain, have a major influence on the reaction rate \cite{poinsot_theoretical_2011}. In this work, we propose to include the topological information by means of spatial convolutions using a \ac{CNN}. 
The CNN is found to perform better when the equivalence ratio $\widetilde{\phi}$ is used instead of the mixture fraction $\widetilde{\xi}$. The filtered equivalence ratio can be approximated as
\begin{equation}
    \widetilde{\phi} = \frac{\widetilde{\xi} \left(1-\xi_s \right)}{\xi_s \left( 1 - \widetilde{\xi} \right)} \, \text{.}
    \label{eq:phi_tilde}
\end{equation}
The \ac{CNN} takes as input the field of $\widetilde{c}$ and $\widetilde{\phi}$ over a domain $\Omega$, performs successive convolutions and operations, and outputs an approximation of $\overline{\dot{\omega}}$ over the same domain $\Omega$. 
The function $f_\mathrm{CNN}$ therefore reads
\begin{equation}
    \boldsymbol{\overline{\dot{\omega}}^\mathrm{NN}} = f_\mathrm{CNN} \left( \boldsymbol{\widetilde{c}}, \boldsymbol{\widetilde{\phi}} \right) \approx \boldsymbol{\overline{\dot{\omega}}} \, \text{.}
    \label{eq:f_CNN}
\end{equation}
\acp{CNN} were designed to learn spatial hierarchies of features, from low- to high-level patterns. They are therefore well fitted for recognizing complex flame topology and learning associated burning rates. Furthermore, the convolutions enable to train models on large inputs via parameter sharing. 
The application of \acp{CNN} to the modeling of flame features has already been successfully performed in Refs~\cite{lapeyre_training_2019,lapeyre_-posteriori_2018,seltz_direct_2019,nikolaou_progress_2019} for fuel-air mixture without significant preferential diffusion effects, unlike the H\textsubscript{2}-air mixtures considered in this paper. 

\begin{figure}[htbp!]%
\FIG{\includegraphics[width=1.\textwidth]{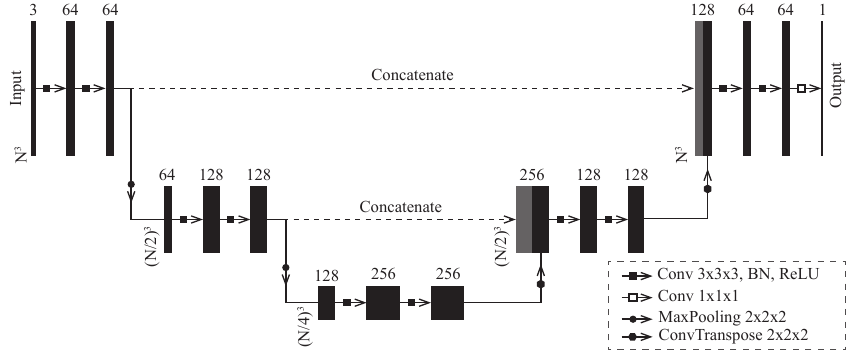}}
{\caption{U-net type architecture used in the present work. Each black box corresponds to a multi-channel feature map. The number of channels is denoted on top of the box. The input sample has three channels: $\widetilde{c}$, $\widetilde{\phi}$ and $\delta_L^0/\delta_L^1$. The ratio $\delta_L^0/\delta_L^1$ is the inverse of the laminar flame thickening due to filtering (Section~\ref{sec:filtering}). The original size of the cubic sample is $\mathrm{N^3}$. It is then reduced to $\mathrm{(N/2)^3}$ and $\mathrm{(N/4)^3}$ during the contracting path before to go back to the original size during the expansive path. Gray boxes represent copied feature maps. The arrows denote the different operations}
\label{fig:CNN_structure}}
\end{figure}

The CNN architecture used is inspired by the U-net architecture originally developed for image segmentation \cite{Ronneberger} (Fig.~\ref{fig:CNN_structure}). However, the output activation function is replaced by a linear activation function for the regression task considered in this work. A similar architecture for the approximation of flame surface density has already been used successfully \cite{lapeyre_training_2019,lapeyre_-posteriori_2018}. Using $64$ channels in the first multi-channel feature map instead of $32$ in Refs~\cite{lapeyre_training_2019,lapeyre_-posteriori_2018} increased the accuracy of the approximation of $\overline{\dot{\omega}}$ (Eq.~(\ref{eq:f_CNN})) in the present work. In total, the network comprises $5,424,193$ trainable parameters.

\subsection{Data generation and training strategy}\label{sec:data_training_strategy}

\begin{figure}[htbp!]%
\FIG{\includegraphics[width=1\textwidth]{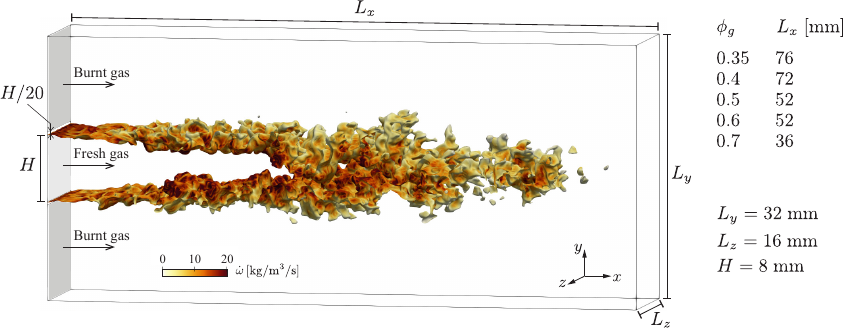}}
{\caption{Diagram of the slot burner configuration used to generate the \ac{DNS} database. The flame is depicted by an iso-surface at a progress variable $c=0.5$ colored by $\dot{\omega}=-\dot{\omega}_\mathrm{H_2}$. The dimensions of the domain are annotated on the right. The length $L_x$ is adapted to the length of the turbulent flame brush, which is function of the global equivalence ratio $\phi_g$}
\label{fig:slot_burner}}
\end{figure}

The data generation strategy consists in filtering and downsampling data from \ac{DNS} to emulate \ac{LES} data with known filtered quantities. The configuration chosen to generate training data is a slot burner (Fig.~\ref{fig:slot_burner}) at constant pressure $P=1$~atm and fresh gas temperature $T^u=300$~K, as in Refs~\cite{lapeyre_training_2019,lapeyre_-posteriori_2018}. The physical domain consists of a central inlet where a premixed H\textsubscript{2}-air mixture flows at a bulk velocity $U_b=24$~$\mathrm{m/s}$ with velocity fluctuation $u^\prime=2.4$~$\mathrm{m/s}$, surrounded by two laminar coflows where burnt gas flows at a bulk velocity $U_c=3.6$~$\mathrm{m/s}$. The injection of turbulence at the central inlet corresponds to homogeneous and isotropic turbulence using a Passot-Pouquet turbulence spectrum \cite{Passot_Pouquet_1987} with an integral length scale $l_t=2$~mm. The domain is rectangular with periodic boundary conditions in the $z$-direction. Adiabatic walls are specified in the $y$-direction. Both inlets and outlet are specified in the $x$-direction. Exact dimensions are annotated on Fig.~\ref{fig:slot_burner}. The configuration is very similar to that used in Refs~\cite{berger_synergistic_2022,coulon_direct_2023,gaucherand_dns_2024} to study H\textsubscript{2}-air and H\textsubscript{2}/NH\textsubscript{3}-air premixed turbulent flames. This configuration is computed for five different equivalence ratios (Table~\ref{tab:cases}) to assess the ability of the \ac{CNN} to approximate the reaction rate over a range of equivalence ratios. All other parameters are kept constant. The Reynolds number of the central inlet is about $10,000$ for all cases. 

\begin{table}[htbp!]
\tabcolsep=0pt%
\TBL{\caption{Global equivalence ratio $\phi_g$ for the five different \ac{DNS} cases. The laminar flame characteristic values are also indicated. $S_L^0$ is the laminar flame speed, $\delta_{th,L}^0$ the laminar thermal flame thickness, $T^{ad}$ the adiabatic flame temperature}\label{tab:cases}}
{
\begin{tabular*}{\textwidth}{@{\extracolsep{\fill}} llllll}
\hline
Case & $\phi_g=0.35$ & $\phi_g=0.4$ & $\phi_g=0.5$ & $\phi_g=0.6$ & $\phi_g=0.7$ \\ \hline
$S_L^0$ [m/s] &  $9.453 \cdot 10^{-2}$    &  $1.925 \cdot 10^{-1}$   &  $4.972 \cdot 10^{-1}$   &  $8.845 \cdot 10^{-1}$   &  $1.302$   \\
$\delta_{th,L}^0$ [m] &  $1.146 \cdot 10^{-3}$    &  $7.176 \cdot 10^{-4}$   &  $4.809 \cdot 10^{-4}$   &  $4.339 \cdot 10^{-4}$   &   $4.171 \cdot 10^{-4}$  \\
$T^{ad}$ [K] &   $1311$   &  $1428$   &  $1646$   &  $1844$   &  $2020$   \\
\hline
\end{tabular*}
}
\end{table}

\begin{figure}[htbp!]%
\FIG{\includegraphics[width=1.\textwidth]{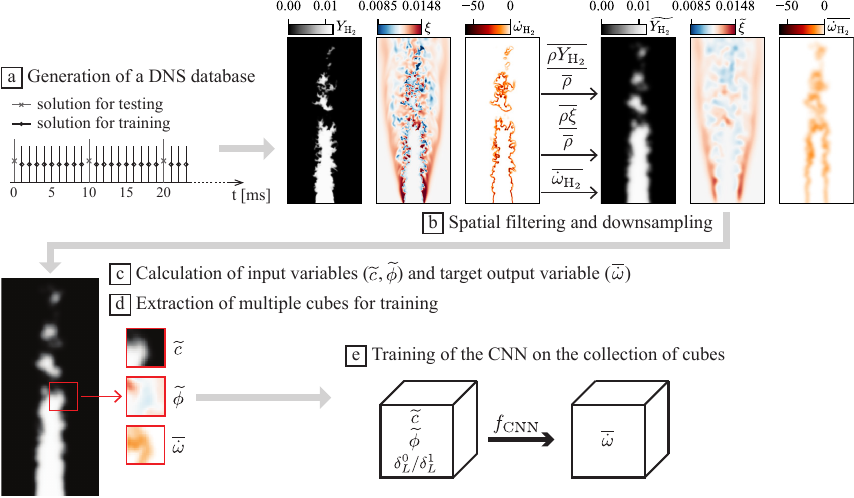}}
{\caption{Diagram of the strategy used to generate data and train the CNN. $\delta_L^0/\delta_L^1$ is the inverse of the laminar flame thickening due to filtering (Section~\ref{sec:filtering}). $\widetilde{\phi}$ is calculated from $\widetilde{\xi}$ using Eq.~(\ref{eq:phi_tilde})}
\label{fig:ML_strategy}}
\end{figure}

The data generation and training strategy is based on processing multiple snapshots from the five \acp{DNS} (Fig.~\ref{fig:ML_strategy}):
\begin{enumerate}[a.]
\item The five DNS cases (Table~\ref{tab:cases}) are computed for $59$~ms from a statistically steady state. \ac{3D} instantaneous solutions are saved every ms. Solutions with an index multiple of 10 are set aside for testing after training. All other solutions are used for training and validation data. This generates 6 solutions for testing and 54 solutions for training and validation data for each DNS case.
\item A spatial filtering is performed on each solution to obtain $\overline{\rho}$, $\overline{\rho Y_\mathrm{H_2}}$, $\overline{\rho \xi}$ and $\overline{\dot{\omega}}_\mathrm{H_2}$. The filtered variables are downsampled to a coarser grid, representing an \ac{LES} grid. The filtering and downsampling operation is described in Section~\ref{sec:filtering}.
\item The filtered progress variable $\widetilde{c}$ is calculated using Eq.~(\ref{eq:c_tilde}) with $\widetilde{Y}_\mathrm{H_2} = \overline{\rho Y_\mathrm{H_2}}/\overline{\rho}$ and $\widetilde{\xi} = \overline{\rho \xi}/\overline{\rho}$. The filtered equivalence ratio $\widetilde{\phi}$ is calculated using Eq.~(\ref{eq:phi_tilde}) with $\widetilde{\xi} = \overline{\rho \xi}/\overline{\rho}$. The filtered burning rate $\overline{\dot{\omega}}$ is calculated using Eq.~(\ref{eq:omega_bar}) with $\overline{\dot{\omega}}_\mathrm{H_2}$.
\item It is important that the \ac{CNN} does not learn the specific configuration of the slot burner, but only learns generic flame elements. Depending on the effective receptive field of the \ac{CNN}, training using the emulated LES solutions directly could encode in the network information on the distance to the inlets, outlet or the wall for example. This would drastically reduce the CNN ability to generalize, and must be avoided. To address this, training is performed on randomly selected cubes in the domain as in Refs~\cite{lapeyre_training_2019,lapeyre_-posteriori_2018}. This also enables the CNN to be invariant to translation \cite{JMLR:v22:21-0019}.
\item The \ac{CNN} is trained to approximate $\overline{\dot{\omega}}$ (Eq.~(\ref{eq:f_CNN})) using data from the cubic samples. 10\% of samples are randomly selected and set aside for validation, the rest are used for training. The complete description of the training is given in Section~\ref{sec:training}.
\end{enumerate}

\subsubsection{Direct numerical simulation}\label{sec:DNS}

\ac{DNS} of the slot burner cases (Table~\ref{tab:cases}) are performed using the AVBP \cite{schonfeld_steady_1999,gicquel_high_2011} massively parallel code solving the compressible multi-species Navier-Stokes equations. A third order accurate Taylor–Galerkin scheme is adopted for discretization of the convective terms \cite{colin_development_2000}. \acp{NSCBC} \cite{poinsot_boundary_1992} are imposed at the inlets (relaxation factor of $1000$ $\mathrm{s^{-1}}$) and at the outlet (relaxation factor of $200$ $\mathrm{s^{-1}}$). 
Dynamic viscosity $\mu$ follows a power law function of temperature $T$
\begin{equation}
    \mu = \mu_{0}\left(\frac{T}{T_{0}}\right)^\gamma \text{ ,}
\end{equation}
with $\mu_{0}=8.062 \times 10^{-5}$ $\mathrm{kg /m /s}$, $T_{0}=2.645 \times 10^{3}$ K and $\gamma=6.481 \times 10^{-1}$. 
Thermal diffusivity is computed from the viscosity using a constant Prandtl number: $\mathrm{Pr} = 0.66$. 
Species diffusivities are computed using a constant Schmidt number specific for each species (Table~\ref{tab:Sc_Le}). This approach takes into account non-unity Lewis numbers and preferential diffusion between the different species. 
It is verified that the errors made by the simplified transport description are negligible by comparing the results with simulations using a mixture-averaged transport model, see supplementary material. 
Soret and Dufour transport processes are ignored in the simulations of the present work. Including these transport processes could enhance thermodiffusive instabilities, increase source terms and flame surface curvature \cite{schlup_validation_2018}. Although this would require another training on the new database, it does not affect the conclusions of this work. Indeed, it has been shown that neglecting these effects still leads to simulations that capture much of the peculiar behaviour of lean premixed H\textsubscript{2}-air flames including thermodiffusive instabilities and enhanced burning rates \cite{schlup_validation_2018,aspden_turbulence-chemistry_2015,aspden_numerical_2017,pitsch_transition_2024}. 
Hydrogen chemical kinetics relies on the San Diego mechanism \cite{saxena_testing_2006}, already successfully used for H\textsubscript{2}-air premixed combustion in Ref.~\cite{coulon_direct_2023}. This mechanism comprises $9$ species and $21$ reactions. 

\begin{table}[htbp!]
\tabcolsep=0pt%
\TBL{\caption{Schmidt $\mathrm{Sc}_k$ and Lewis $\mathrm{Le}_k$ numbers for the species used for the \ac{DNS} of H\textsubscript{2}-air flames}\label{tab:Sc_Le}}
{
\begin{tabular*}{\textwidth}{@{\extracolsep{\fill}}llllllllll}
\hline
Species $k$ & $\mathrm{H_2}$ & $\mathrm{H}$ & $\mathrm{O_2}$ & $\mathrm{OH}$ & $\mathrm{O}$ & $\mathrm{H_2O}$ & $\mathrm{HO_2}$ & $\mathrm{H_2O_2}$ & $\mathrm{N_2}$ \\ \hline
$\mathrm{Sc}_k$ & $0.23$ & $0.14$ & $0.80$ & $0.53$ & $0.52$ & $0.58$ & $0.80$ & $0.81$ & $0.91$ \\
$\mathrm{Le}_k$ & $0.34$ & $0.21$ & $1.20$ & $0.80$ & $0.79$ & $0.88$ & $1.21$ & $1.22$ & $1.37$ \\
\hline
\end{tabular*}
}
\end{table}

The mesh is a homogeneous cartesian grid with constant element size $\Delta_x = 0.1$~mm. This gives a resolution index $\mathrm{RI}=\delta_{th,L}^0/\Delta_x+1$ that varies between $12.5$ and $5.2$ for $\phi_g$ between $0.35$ and $0.7$. $\delta_{th,L}^0$ is the laminar thermal flame thickness. 
A mesh sensitivity study was carried out where we verified that turbulent combustion statistics are not affected by mesh refinements. This study is included as supplementary material. 
The length of the domain in the $x$-direction $L_x$ is adapted to the length of turbulent the flame brush. It varies from $76$~mm for $\phi_g = 0.35$ to $36$~mm for $\phi_g = 0.7$ (Fig.~\ref{fig:slot_burner}). The number of grid points accordingly varies from $39$ to $19$ million. Calculations were carried out on the new Alps Research Infrastructure at the Swiss National Supercomputing Centre (CSCS) using AMD EPYC 7742 CPUs. A total of $1.6$ million core-hours were used to generate the database. Depending on the grid size, computations are performed using between $5,120$ and $2,560$ computing cores with good efficiency thanks to the massively parallel computing ability of the AVBP code.

\subsubsection{Filtering and downsampling}\label{sec:filtering}

A spatial filtering is performed to obtain $\overline{\rho}$, $\overline{\rho Y_\mathrm{H_2}}$, $\overline{\rho \xi}$ and $\overline{\dot{\omega}_\mathrm{H_2}}$, in order to calculate $\widetilde{c}$, $\widetilde{\phi}$ and $\overline{\dot{\omega}}$ (Eqs~(\ref{eq:c_tilde}), (\ref{eq:phi_tilde}) and (\ref{eq:omega_bar})). The filtered quantity of a variable $\varphi$ from the \ac{DNS} field is defined by Eq.~(\ref{eq:filtering}). 
The filtered \ac{3D} fields are obtained by successively filtering on each direction $x$, $y$ and $z$ using a \ac{1D} Gaussian filter. The \ac{1D} filter function is written in discrete form as
\begin{equation}
    F(u)= \begin{cases}e^{-\frac{1}{2}\left(\frac{u}{\sigma}\right)^2} & \text { if } u \in \left[ -z \; ..  +z \right] \\ 0 & \text { otherwise }\end{cases} \, \text{with} \, z = 4 \sigma \, \text{,}
    \label{eq:gaussian_filter}
\end{equation}
and then normalized by its sum $\sum_{u=-z}^z F(u)$. The filtered fields are then downsampled to a coarser grid to emulate \ac{LES} fields. Three different standard deviations $\sigma$ are used in this work to emulate three different \ac{LES} resolutions (Table~\ref{tab:LES_parameters}). For each $\sigma$, a downsampling factor DSF is associated. It is chosen so as to have a \ac{LES} grid with a resolution index $\mathrm{RI}=\delta_{c,L}^1/\Delta_x+1$ of about 6. $\delta_{c,L}^0$ and $\delta_{c,L}^1$ are the laminar flame thickness based on the progress variable of a non-filtered and filtered \ac{1D} flame
\begin{equation}
    \delta_{c,L}^0 = \left( \mathrm{max} \left| \nabla c \right| \right) ^ {-1} \, \text{and} \, \delta_{c,L}^1 = \left( \mathrm{max} \left| \nabla \widetilde{c} \right| \right) ^ {-1} \, \text{.}
    \label{eq:dl01c}
\end{equation}

\begin{table}[htbp!]
\tabcolsep=0pt%
\TBL{\caption{Flame characteristics for the three sets of LES parameters ($\sigma$, DSF) used in this work. $\Delta_x$ is the grid cell size, equals to DSF times the original cell size of $0.1$~mm. $\delta_{c,L}^0$ and $\delta_{c,L}^1$ are the laminar flame thickness based on the progress variable of a non-filtered and filtered \ac{1D} flame (Eq.~(\ref{eq:dl01c})). $\mathrm{RI}=\delta_{c,L}^1/\Delta_x+1$ is the resolution index of the filtered flame on the LES grid}\label{tab:LES_parameters}}
{
\begin{tabular*}{\textwidth}{@{\extracolsep{\fill}} lllllll}
\cline{2-7}
                                                                                       & Case    & $\phi_g=0.35$ & $\phi_g=0.4$ & $\phi_g=0.5$ & $\phi_g=0.6$ & $\phi_g=0.7$ \\ \cline{2-7} 
\multirow{3}{*}{\begin{tabular}[c]{@{}l@{}}$\sigma=4$\\ $\mathrm{DSF}=2$\\ $\Delta_x=0.2$~mm\end{tabular}} & $\delta_{c,L}^1$~[m]     &   $1.41 \cdot 10^{-3}$   &  $1.06 \cdot 10^{-3}$   &  $9.28 \cdot 10^{-4}$   &  $9.01 \cdot 10^{-4}$   &   $9.09 \cdot 10^{-4}$  \\
                                                                                       & $\delta_{c,L}^0/\delta_{c,L}^1$ & $0.81$     &  $0.59$   &   $0.41$  &   $0.36$  &  $0.33$   \\
                                                                                       & RI      &   $8.03$   &  $6.30$   &  $5.64$   &   $5.51$  &  $5.55$  \\
\cline{2-7}
\multirow{3}{*}{\begin{tabular}[c]{@{}l@{}}$\sigma=8$\\ $\mathrm{DSF}=4$\\ $\Delta_x=0.4$~mm\end{tabular}} & $\delta_{c,L}^1$~[m]    &   $2.06 \cdot 10^{-3}$   &  $1.83 \cdot 10^{-3}$  &   $1.75 \cdot 10^{-3}$  &  $1.75 \cdot 10^{-3}$   &   $1.75 \cdot 10^{-3}$  \\
                                                                                       & $\delta_{c,L}^0/\delta_{c,L}^1$ & 0.55     &   0.34  &   0.21  &  0.18   &   0.17  \\
                                                                                       & RI      &   $6.16$   &   $5.58$  &  $5.38$    &   $5.36$  &    $5.37$  \\
\cline{2-7}
\multirow{3}{*}{\begin{tabular}[c]{@{}l@{}}$\sigma=16$\\ $\mathrm{DSF}=8$\\ $\Delta_x=0.8$~mm\end{tabular}} & $\delta_{c,L}^1$~[m]    &  $3.66 \cdot 10^{-3}$    &   $3.53 \cdot 10^{-3}$  &  $3.47 \cdot 10^{-3}$   &  $3.45 \cdot 10^{-3}$   &   $3.45 \cdot 10^{-3}$  \\
                                                                                       & $\delta_{c,L}^0/\delta_{c,L}^1$ & $0.31$     &   $0.18$  &   $0.11$  &   $0.09$  &  $0.09 $  \\
                                                                                       & RI      &   $5.58$  &   $5.41$  &  $5.34$   &  $5.32$   &   $5.31$  \\
\cline{2-7}
\end{tabular*}
}
\end{table}

\subsubsection{Training method}\label{sec:training}

The generated cubic samples (Fig.~\ref{fig:ML_strategy}) of size $16 \times 16 \times 16$ points are used to train the \ac{CNN} to approximate $\overline{\dot{\omega}}$ (Eq.~(\ref{eq:f_CNN})). A set of 10\% of the samples is randomly selected to form a validation dataset and monitor overfitting. The remaining samples are used for training. At each epoch, each cubic sample in the training dataset has a $25$\% probability of undergoing between one and three $90^{\circ}$ rotations along any axis, or of being flipped in any direction. It is done to add variability to the orientation of the flame front in the dataset. The model must have no preferential orientation and learn an isotropic function. 

Training is accomplished by backpropagation. The trainable parameters of the \ac{CNN} are updated iteratively on batches of data from the training dataset by stochastic gradient descent with the AdamW optimizer \cite{kingma_adam_2017,loshchilov_decoupled_2019}. The parameters are adjusted to minimize the \ac{MSE} loss function over all output pixels of the prediction compared to the target. The batch size per GPU is $200$ and the training is terminated after $1,500$ epochs. The base learning rate equals $10^{-3}$. 
The evolution of the loss function on the training and validation datasets during training is given in Section~\ref{sec:results}.

The PyTorch \cite{ansel_pytorch_2024} Python library is used to implement and train the \ac{CNN}. The \ac{DDP} feature in PyTorch is employed to parallelize the training process across four NVIDIA H100 GPUs. Each GPU performs its own forward and backward pass on a subset of the training dataset. The gradients computed on each GPU are then synchronized and averaged to update the model parameters. Training performances are reported in Section~\ref{sec:results}.

\FloatBarrier
%
\section{Results and discussion}\label{sec:results}

\subsection{Training and testing}
\begin{figure}
\centering
\begin{minipage}{.5\textwidth}
  \centering
  \includegraphics[width=1\linewidth]{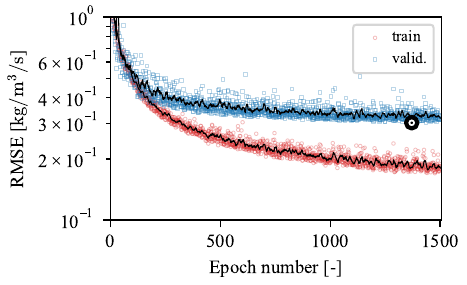}
\end{minipage}%
\begin{minipage}{.5\textwidth}
  \centering
  \includegraphics[width=1\linewidth]{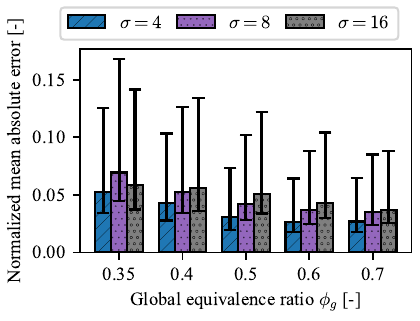}
\end{minipage}
\caption{Left: Evolution of the \acf{RMSE} during training, evaluated over the training dataset (red circles) and the validation dataset (blue squares). Black solid lines are moving averages of the RMSE. The black circle shows the lowest RMSE over the validation dataset. The model parameters at this specific epoch are selected. Right: Normalized mean absolute error over the testing solutions (Eq.~(\ref{eq:NMAE})) for the different equivalence ratios and LES parameters used for building the training dataset. Error bars show first and third quartiles of the data points}
\label{fig:loss_training_MAE_testing}
\end{figure}

\begin{figure}[htbp!]%
\FIG{\includegraphics[width=0.9\textwidth]{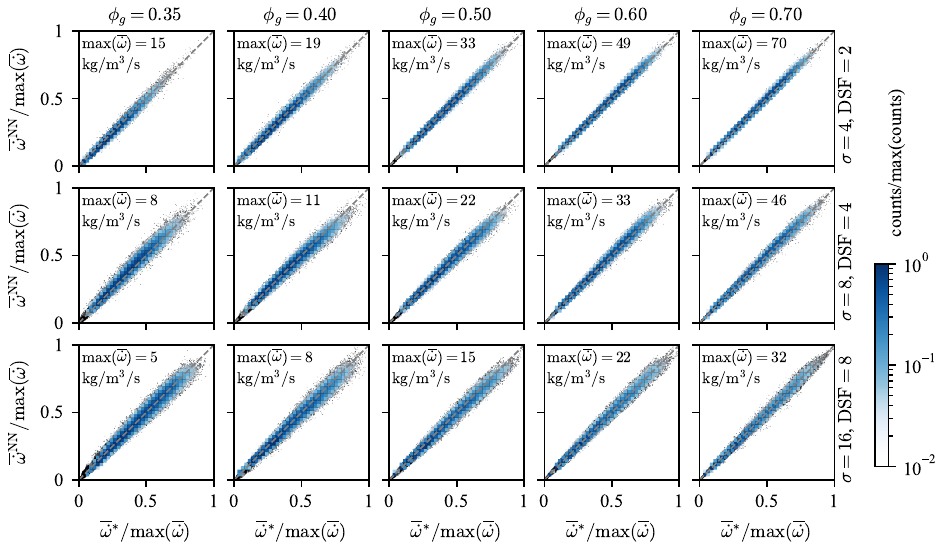}}
{\caption{Scatter plots with 2D histograms: CNN-modeled burning rate $\overline{\dot{\omega}}^{\mathrm{NN}}$ versus ground-truth filtered burning rate $\overline{\dot{\omega}}^{\ast}$. Individual values are normalized by the maximum burning rate in the datasets. The points used for the histograms have a progress variable $c$: $0.05\leq c \leq 0.95$. Histogram values below the colour scale are transparent. Gray dashed line indicates $x=y$ (i.e. zero error). Each column corresponds to a global equivalence ratio. Each row corresponds to a set of LES parameters (filtering and downsampling). Data are collected from the testing solutions}
\label{fig:scatter_eqratios_filtersizes}}
\end{figure}

The training and validation database comprises $40 \times 54 \times 5 \times 3 = 32,400$ cubic samples ($40$ random cubes per solution, $54$ solutions per case, $5$ different equivalence ratios, $3$ different sets of LES parameters (filtering and downsampling)). $3,240$ ($10$\%) of these cubes are randomly set aside for validation. 
Training the CNN model for $1,500$ epochs on this database takes $47$ minutes. $1,500$ epochs are sufficient because the loss function evaluated on the validation dataset no longer decreases significantly (Fig.~\ref{fig:loss_training_MAE_testing}). The set of model parameters giving rise to the lowest loss is selected as the best model (black circle in Fig.~\ref{fig:loss_training_MAE_testing}). This best model is used to assess the ability of the CNN to approximate burning rates on the test solutions (full domain, never seen during training). The Normalized Mean Absolute Error (NMAE) is used to estimate the accuracy of the CNN model (Fig.~\ref{fig:loss_training_MAE_testing}). It is defined as:
\begin{equation}
    \mathrm{NMAE} = \frac{1}{\mathrm{mean}\left(\overline{\dot{\omega}}^{\ast}\right)} \frac{1}{N} \sum_{i=1}^N \left| \overline{\dot{\omega}}^{\ast}_i - \overline{\dot{\omega}}^{\mathrm{NN}}_i \right| \, \text{,}
    \label{eq:NMAE}
\end{equation}
where $\overline{\dot{\omega}}^{\ast}$ is the filtered burning rate from the DNS dataset and $\overline{\dot{\omega}}^{\mathrm{NN}}$ is the filtered burning rate modeled by the CNN. The mean absolute error is only about $5$\% of the average burning rate for all cases. The very high accuracy of the CNN model is confirmed by the distribution of CNN-modeled burning rate $\overline{\dot{\omega}}^{\mathrm{NN}}$ versus ground-truth filtered burning rate $\overline{\dot{\omega}}^{\ast}$ (Fig.~\ref{fig:scatter_eqratios_filtersizes}). The log-normalized \ac{2D} histograms show a large majority of points on the $x=y$ axis (i.e. perfect agreement). 
Figure~\ref{fig:scatter_eqratios_filtersizes} shows a visualization of the result using a planar cut through a test solution for the case $\phi_g = 0.4$ and for the three sets of LES parameters $\left( \sigma , \mathrm{DSF} \right)$. One can see the ability of the CNN to retrieve the extremely complex turbulent flame topology for all filter sizes. Even the burning pockets detached from the main flame brush are correctly modeled by the CNN. This is true for all other equivalence ratios. The visualization of the cases $\phi_g = 0.35$, $0.5$, $0.6$ and $0.7$ are given as supplementary material to avoid overloading the manuscript.

\begin{figure}[htbp!]%
\FIG{\includegraphics[width=0.9\textwidth]{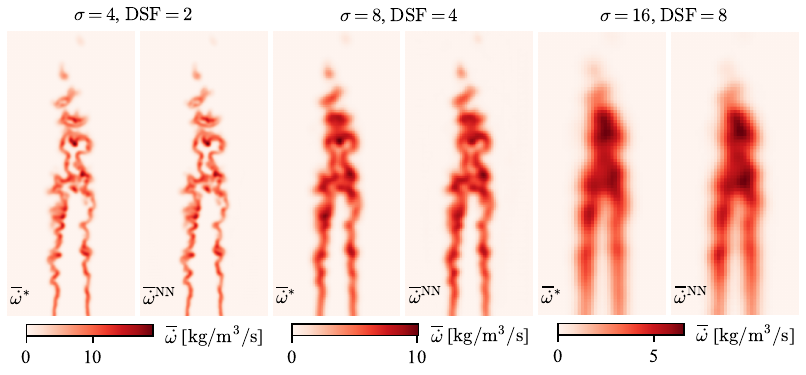}}
{\caption{Planar cut normal to the $z$-axis, in the middle of the domain, colored by the ground-truth filtered burning rate $\overline{\dot{\omega}}^{\ast}$ and the CNN-modeled burning rate $\overline{\dot{\omega}}^{\mathrm{NN}}$ for three sets of LES parameters. The global equivalence ratio is $\phi_g = 0.4$. The complex turbulent flame topology of the burning rates is remarkably well reproduced by the CNN}
\label{fig:cut_omega_phi04}}
\end{figure}

\subsection{Generalization ability}

The ability of the CNN model to generalize outside the cases for which it has been trained is now assessed. For this, two new sets of LES parameters $\left(\sigma=6, \mathrm{DSF}=3\right)$ and $\left(\sigma=12, \mathrm{DSF}=6\right)$, and two new equivalence ratios $\phi_g=0.45$ and $0.55$ are used. 

\begin{figure}[htbp!]%
\FIG{\includegraphics[width=0.9\textwidth]{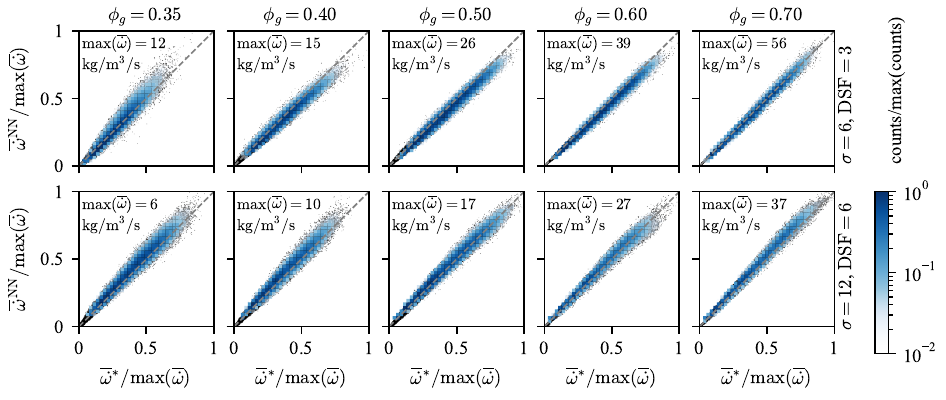}}
{\caption{Scatter plots with 2D histograms: CNN-modeled burning rate $\overline{\dot{\omega}}^{\mathrm{NN}}$ versus ground-truth filtered burning rate $\overline{\dot{\omega}}^{\ast}$. Individual values are normalized by the maximum burning rate in the datasets. The points used for the histograms have a progress variable $c$: $0.05\leq c \leq 0.95$. Histogram values below the colour scale are transparent. Gray dashed line indicates $x=y$ (i.e. zero error). Each column corresponds to a global equivalence ratio. Each row corresponds to a set of LES parameters (filtering and downsampling). Data are collected from solutions with two sets of LES parameters that were not included into the training dataset. The CNN approximates the burning rates with good accuracy, demonstrating the ability to generalize to other LES parameters from which it has been trained}
\label{fig:scatter_generalization_filtersizes}}
\end{figure}

Filtering and downsampling with the two new sets of LES parameters are performed on test solutions of the existing DNS with global equivalence ratios $\phi_g = 0.35$, $0.4$, $0.5$, $0.6$ and $0.7$. Although the distribution of points $\overline{\dot{\omega}}^{\mathrm{NN}}$ versus $\overline{\dot{\omega}}^{\ast}$ is a little more scattered (Fig.~\ref{fig:scatter_generalization_filtersizes}), the majority of the CNN-modeled burning rates are in very good agreement with their ground-truth counterparts. This demonstrates the ability of the model to generalize between two LES filter sizes for which it has been trained. It is important to note, however, that training with all three filter sizes (Tab.~\ref{tab:LES_parameters}) is necessary to achieve this result. We have tried to train the CNN with only two filter sizes: $\sigma=4$ and $16$, and found that the resulting model has a significant bias for $\sigma=8$, see supplementary material.

\begin{figure}[htbp!]%
\FIG{\includegraphics[width=0.9\textwidth]{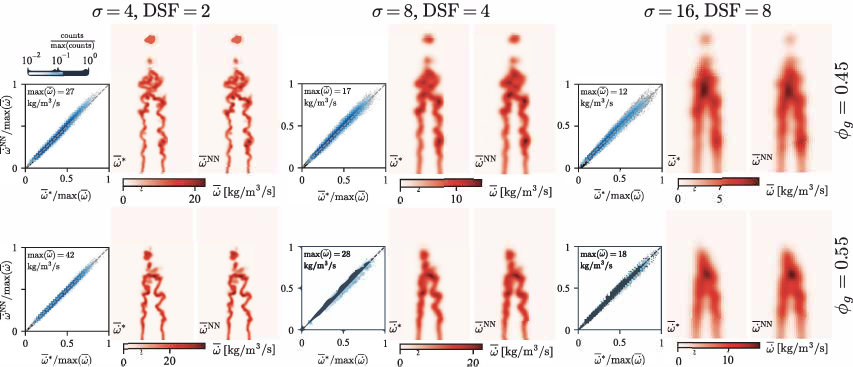}}
{\caption{Scatter plots and 2D histograms together with planar cuts comparing CNN-modeled burning rate $\overline{\dot{\omega}}^{\mathrm{NN}}$ to ground-truth filtered burning rate $\overline{\dot{\omega}}^{\ast}$. See caption of Fig.~\ref{fig:scatter_eqratios_filtersizes} for a full description of how the histograms are constructed. The two global equivalence ratios $\phi_g=0.45$ and $0.55$ were not included in the training dataset. The CNN approximates the burning rates with high accuracy, demonstrating the ability to generalize to other equivalence ratios from which it has been trained}
\label{fig:scatter_cut_generalization_eqratios}}
\end{figure}

The ability of the model to approximate burning rates for other global equivalence ratios is now assessed. For this, two new DNS are performed with a global equivalence ratio $\phi_g = 0.45$ and $0.55$ and six solutions per DNS (snapshot every $1$~ms) are used to evaluate the model on the filter sizes $\sigma=4$, $8$ and $16$ (Fig.~\ref{fig:scatter_cut_generalization_eqratios}). The distribution of points $\overline{\dot{\omega}}^{\mathrm{NN}}$ versus $\overline{\dot{\omega}}^{\ast}$ is a little more scattered compared to inference on solutions with a global equivalence ratios used for training. The majority of the points are nevertheless clustered along the $x=y$ axis: the approximation of the burning rates is still very good. The planar cuts show a flame topology very well retrieved by the CNN model. These excellent results are due to the large number of equivalence ratios included in the training dataset. We have trained the CNN with only $\phi_g = 0.35$, $0.4$, $0.6$ and $0.7$ (i.e. without $\phi_g = 0.5$) and that model is flawed for the global equivalence ratio $\phi_g = 0.5$, see supplementary material.

These generalization results are very promising and show that a CNN model can generalize with a reasonable amount of training cases. This is particularly important with regard to the global equivalence ratio $\phi_g$. Computing a DNS for each global equivalence ratio has a significant computing cost. In contrast, the computational cost of a filtering and downsampling operation to produce a new set of LES parameters $\left( \sigma , \mathrm{DSF} \right)$ is negligible.

\subsection{Comparison with a filtered tabulated chemistry approach}

A filtered tabulated chemistry modeling is implemented to compare CNN results with an established approach for modeling filtered burning rates. In the filtered tabulated chemistry framework, the flame structure in the direction normal to the flame front is assumed identical to the structure of a planar \ac{1D} freely propagating premixed laminar flame \cite{fiorina_filtered_2010}. 
For a given filter size, it is then possible to apply the filter operation introduced in Section~\ref{sec:filtering} (Eqs~(\ref{eq:filtering}) and (\ref{eq:gaussian_filter})) to a precomputed 1D flame at a given global equivalence ratio $\phi_g$ and extract the burning rates directly as
\begin{equation}
    \overline{\dot{\omega}} \approx \overline{\dot{\omega}}^{\,+} \left(\widetilde{c}, \phi_g \right) \, \text{,}
\end{equation}
where $\overline{\dot{\omega}}^{\,+}$ is the filtered burning rate of the 1D flame, mapped as a function of the filtered progress variable $\widetilde{c}$ for a given filter size $\sigma$. 
To account for the local fluctuation of the equivalence ratio in the domain, it is possible to retrieve the reaction rate in a 1D flame computed with the local equivalence ratio $\phi$ used to define the fresh gas composition. In this case, we write
\begin{equation}
    \overline{\dot{\omega}} \approx \overline{\dot{\omega}}^{\,+} \left(\widetilde{c}, \widetilde{\phi} \right) \, \text{.}
\end{equation}
Flame wrinkling at the subfilter-scale level is lost in \ac{LES}. A subfilter-scale wrinkling factor $\Xi$ needs to be applied to compensate for the loss in flame surface \cite{poinsot_theoretical_2011}. The filtered burning rate for filtered tabulated chemistry modeling therefore reads
\begin{equation}
    \overline{\dot{\omega}}^{F} = \Xi \, \overline{\dot{\omega}}^{\,+} \left(\widetilde{c}, \phi_g \right) \; \text{or} \; \overline{\dot{\omega}}^{FC} = \Xi \, \overline{\dot{\omega}}^{\,+} \left(\widetilde{c}, \widetilde{\phi} \right) \; \text{,}
    \label{eq:omegaFC}
\end{equation}
depending on whether the global or local equivalence ratio is used to define the fresh gas composition in the corresponding 1D flame. 
The wrinkling factor $\Xi$ is usually modeled by algebraic expressions. Models based on a fractal description of the flame front \cite{gouldin_application_1987,gouldin_chemical_1989} are very common. They assume that in a range of physical scales bounded by an inner cutoff $\eta$ and an outer cutoff $L$, the flame front is a fractal surface of dimension $2 \leq D_f \leq 3$. The wrinkling factor is given by
\begin{equation}
    \Xi = \left(\frac{L}{\eta}\right)^{D_f-2} \, \text{.}
    \label{eq:Xi}
\end{equation}
$L$ corresponds to the size of the largest unresolved wrinkles, which is of the order of the thickness of the LES flame estimated as $\delta_{c,L}^1$ (Eq.~(\ref{eq:dl01c})). $\eta$ is the size of the smallest wrinkles and varies from $\eta=L$ (i.e. no unresolved wrinkling, $\Xi=1$) to $\eta=\delta_{c,L}^0$ (i.e. filling of the space by a flame surface of fractal dimension $D_f$ between cut-off scales $\delta_{c,L}^0$ and $L$, $\Xi=\left( L / \delta_{c,L}^0 \right)^{D_f-2}$). The latter situation is reached when turbulence intensity is sufficiently high, which is often the case in practice \cite{veynante_analysis_2012,veynante_analysis_2015}. We then assume $\eta=\delta_{c,L}^0$ in Eq.~(\ref{eq:Xi}), maximizing the flame surface density at the subfilter-scale level. The fractal dimension is estimated as $D_f=2.5$ following the work of Charlette et al. \cite{charlette_power-law_2002}. 
The set of 1D flames are computed using the San Diego mechanism \cite{saxena_testing_2006} for the chemical kinetics (same as the training database for the CNN) and a mixture-averaged transport model for the diffusivities. 

\begin{figure}[htbp!]%
\FIG{\includegraphics[width=0.9\textwidth]{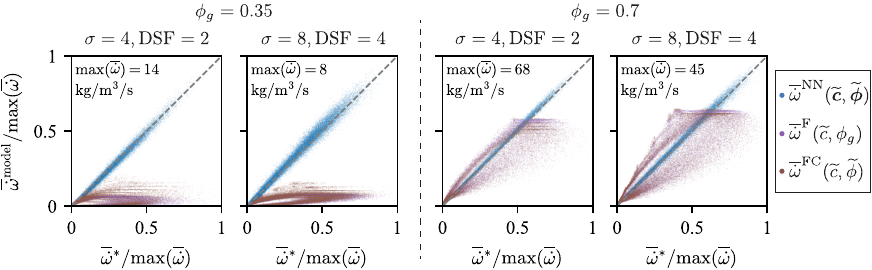}}
{\caption{Scatter plots of the CNN-modeled burning rate $\overline{\dot{\omega}}^{\mathrm{NN}}$ and the filtered tabulated chemistry burning rate $\overline{\dot{\omega}}^{\mathrm{F}}$ or $\overline{\dot{\omega}}^{\mathrm{FC}}$ (Eq.~(\ref{eq:omegaFC})) versus ground-truth filtered burning rate $\overline{\dot{\omega}}^{\ast}$. Individual values are normalized by the maximum burning rate in the datasets. Gray dashed line indicates $x=y$ (i.e. zero error). Two cases of LES parameter sets are presented for two global equivalence ratios: the lowest equivalence ratio $\phi_g=0.35$ for which the filtered tabulated chemistry is significantly flawed due to very strong thermodiffusive effects; the highest equivalence ratio $\phi_g=0.7$ for which thermodiffusive effects are less important}
\label{fig:scatter_FTAC_CNN}}
\end{figure}

CNN-based modeling significantly outperforms the filtered tabulated chemistry one, especially for the lean cases (e.g. $\phi_g=0.35$ in Fig.~\ref{fig:scatter_FTAC_CNN}). Due to the extremely high thermodiffusive effects in the very lean cases, the burning rates of the turbulent \ac{3D} field can be much higher than those in a freely propagating planar flame at the same equivalence ratio. For example, for the case $\phi_g=0.35$, $\sigma=4$, $\mathrm{DSF}=2$, the maximum burning rate in the \ac{3D} field $\mathrm{max} ( \overline{\dot{\omega}}^{\ast} )$ is about $17$ times greater than that in a laminar planar flame $\mathrm{max} ( \overline{\dot{\omega}}^{\,+} )$. 
This is due to the fact that flame strain is not considered in the tabulation method applied, and that the subfilter wrinkling factor $\Xi$ only takes into account turbulence-induced wrinkling, ignoring that induced by thermodiffusive instabilities and their interactions with turbulence \cite{berger_synergistic_2022}. 
This $\phi_g=0.35$ case is a truly extreme test case to illustrate the strength of the CNN model. Tabulated chemistry modeling does not fail so spectacularly for equivalence ratios close to stoichiometry, where preferential and differential diffusion effects are less pronounced \cite{berger_intrinsic_2022-1} (e.g. $\phi_g=0.7$ in Fig.~\ref{fig:scatter_FTAC_CNN}). The CNN results are still much more accurate thanks to the consideration of the \ac{3D} flame structure via successive convolutions. 
To further demonstrate the CNN ability to learn from flame topology, training has been performed without information on the equivalence ratio for the case $\phi_g=0.35$. This study is given as supplementary material. It is found that burning rate approximation is still very accurate, proving our initial claim that a CNN is capable of recognizing complex flame topology and inferring associated burning rates by learning spatial hierarchies of features.

%
\section{Conclusion and future work}

Trustworthy LES of lean premixed hydrogen combustion requires advanced modeling methods considering thermodiffusive effects on burning rates at the subfilter-scale level. These effects are extremely complex and their coupling with turbulence is controversial, making the derivation of analytical models very difficult. We therefore propose a pure ML approach where a CNN is trained to approximate LES burning rates from emulated LES fields, generated by explicitly filtering many DNS snapshots. 
The configuration used to generate the training data is a slot burner where a turbulent flow of premixed H\textsubscript{2}-air mixture is placed in between two coflows of burnt gas burnt at the same equivalence ratio. The training dataset is built up by performing five DNSs of this configuration, each with a unique global equivalence ratio. The DNS solutions are then filtered and downsampled using three different sizes to emulate LES solutions. Many small cubes are extracted from the solutions to train the CNN model so that it does not learn the specific configuration but rather generic flame elements, improving its generalization ability later on. 

The trained CNN model retrieves burning rates with very high accuracy on the 3D full-scale test solutions. It is able to model burning rates in a very complex turbulent flame topology with flame elements detached from the main flame brush. In addition, the CNN approximates burning rates with low errors for filter and downsampling parameters and global equivalence ratios other than those for which it has been trained. The ML-based modeling outperforms a priori a classical filtered tabulated chemistry approach with correction for flame surface density due to subfilter-scale flame wrinkling. This is particularly true for very lean flames, where differential diffusion effects are major and undermine modeling based on flame surface density. 

These results are extremely promising. They demonstrate that it is possible to build an ML-based model of the burning rates of turbulent lean premixed H\textsubscript{2}-air flames\textemdash an extremely challenging task\textemdash that can generalize outside the scope for which it has been trained. Generalization is very important if we are really going to consider implementing this type of model in engineering simulation software. This work paves the way for a new modeling approach to turbulent H\textsubscript{2} combustion, which we want to pursue with the scientific community in the very near future. 

We can then draw up a roadmap for integration into an LES code. 
First, the modeling of the subfilter-scale fluxes in Eq.~(\ref{eq:LES_transport_varphi}) remains to be addressed. It could be done through gradient transport models with a turbulent viscosity hypothesis \cite{boger_direct_1998}. If this closure fails to retrieve correct flame propagation speeds, we could consider training the CNN to approximate the fluxes using several output channels. 
Second, the evolution of the entire chemical system must be described. This can be done by training the CNN on the source term of a progress variable, transported instead of the individual species that make up the reactive mixture. This is done, for example, in Refs~\cite{lapenna_data-driven_2021,lapenna_-posteriori_2024} with a tabulation method for the closure of the transport equation of the progress variable. However, the transport of the mixture fraction must be modeled to account for equivalence ratio fluctuations. 
We could also train the model again on a database generated with a single global chemical reaction \cite{millan-merino_new_2024,schiavone_arrhenius-based_2024}. This has the advantage of being able to transport physical species instead of a progress variable, accounting for preferential diffusion through the multi-species Navier-Stokes equations directly. 

Second, there are a number of practical integration challenges to be tackled. In a massively parallel HPC code, the fluid domain is decomposed in many partitions distributed over the MPI tasks (typically one MPI per GPU). When inference is performed on each MPI, artefacts may appear at the boundaries of the partitions~\cite{serhani_graph_2024}. To avoid this, the inference must be given its neighboring context to produce overlapping predictions. 
However, this incurs additional MPI communication costs that must be evaluated on a case by case basis. Another major challenge is related to the voxel grid requirement for the CNN. If the CFD code uses an irregular grid that does not match the voxel/pixel structure, two grids have to be used: one for the fluid solver, one for the CNN inference. This requires the development of efficient methods for interpolating on-the-fly and transferring data between the two grids. The PhyDLL\footnote{\url{https://phydll.readthedocs.io}}~\cite{serhani_graph_2024} open-source library is one effort that could help to address these issues. These questions will be the focus of a future project, now that we have shown that CNN-based modeling of the burning rates is possible with this work.

\begin{Backmatter}

\paragraph{Acknowledgments}
The authors gratefully acknowledge CERFACS for providing the LES solver AVBP and the antares and ARCANE libraries.

\paragraph{Funding Statement}
This research was funded in part by the Swiss National Science Foundation (SNSF) under grant agreement No. 219938. 
This work was supported by a grant from the Swiss National Supercomputing Centre (CSCS) under project ID s1262.

\paragraph{Competing Interests}
The authors declare none.

\paragraph{Data Availability Statement}
Code for training and inference is available via GitLab at \url{https://gitlab.com/male.quentin/cnn_h2flame}. 
A set of solutions from \ac{DNS} is distributed via Kaggle (\url{https://www.kaggle.com}) with the following identifiers:
\begin{itemize}
    \item malquentin/premixed-flame-slot-burner-dns-h2air-phi035
    \item malquentin/premixed-flame-slot-burner-dns-h2air-phi04
    \item malquentin/premixed-flame-slot-burner-dns-h2air-phi05
    \item malquentin/premixed-flame-slot-burner-dns-h2air-phi06
    \item malquentin/premixed-flame-slot-burner-dns-h2air-phi07
\end{itemize}
The DNS database will be integrated into BLASTNet (\url{https://blastnet.github.io/index.html}).

\paragraph{Ethical Standards}
The research meets all ethical guidelines, including adherence to the legal requirements of the study country.

\paragraph{Author Contributions}
Q.M.: conceived the research project, performed the DNS, generated the database, trained the CNN, post-processed the results, performed the analysis, discussed the results, wrote the paper. C.J.L.: discussed the results, reviewed and edited the paper. N.N.: discussed the results, reviewed and edited the paper.

\bibliographystyle{apalike}
\bibliography{zotero,custom}

\end{Backmatter}

\end{document}